# Influencia del potencial de polarización en la deposición de películas delgadas de NiO

Javier García Molleja[a,b], Bruna Regalado[a], Julien Keraudy[b,c], Graciela Salum[a,d], Pierre-Yves Jouan[b]

[a] Yachay Tech University, School of Physical Sciences and Nanotechnology, 100119-Urcuquí, Ecuador
[b] Institut des Matériaux Jean Rouxel (Université de Nantes-CNRS), 2 rue de la Houssinière, 44322 Nantes, Francia
[c] Institut de Recherche Technologique (IRT), Chemin du Chaffault, 44340 Bougenais, Francia
[d] Instituto de Física Rosario (UNR-Conicet), Bv. 27 de Febrero 210 bis, S2000EZP Rosario, Argentina
jgarcia@yachaytech.edu.ec, bregalado@yachaytech.edu.ec, julien.keraudy@cnrs-imn.fr, gsalum@yachaytech.edu.ec,
pierre-yves.jouan@cnrs-imn.fr

*Resumen*— El óxido de níquel (NiO) es un compuesto binario de gran versatilidad en las aplicaciones tecnológicas actuales. Mediante el uso de sputtering por magnetrón en modo reactivo se depositaron películas de NiO bajo diferentes valores de polarización del sustrato de vidrio (0, 50, 100, 200, 300, 400 y 500 V). Dichas películas fueron caracterizadas mediante profilometría, difracción de rayos X, composición elemental y resistividad eléctrica tanto a temperatura ambiente como a bajas temperaturas. Este trabajo determinó que a pesar del carácter aislante del sustrato la tensión residual de las películas decayó con la aplicación de polarización, mientras la texturación en el plano (111) se vio fortalecida. La resistividad eléctrica bajó con el aumento de polarización y a bajas temperaturas se encuentra una activación térmica del proceso de conducción por huecos.

*Palabras Claves*— Óxido de níquel, polarización eléctrica, resistividad eléctrica, tensiones residuales.

*Abstract*— Nickel oxide (NiO) is a binary compound with a lot of applications in the present technology. NiO thin films were deposited by reactive sputtering magnetron under several voltage biases applied in the glass substrates (0, 50, 100, 200, 300, 400 and 500 V). Films were characterized by profilometry, X-ray diffraction, elemental composition and electrical resistivity at room temperature and at low temperatures. The present work showed that despite the insulating behavior of substrate the residual stress was reduced at high biases and the (111) texture was promoted. Electrical resistivity was reduced at high bias and at low temperatures thermal activation of *p*-type conduction was detected.

## I. INTRODUCCIÓN

El óxido de níquel (NiO) es un compuesto binario cerámico con excelentes propiedades desde el punto de vista tecnológico. Si bien su estructura cristalina es romboédrica [1] se puede aproximar su estudio a una estructura fcc [2]. El NiO no es estequiométrico [3] y se comporta como un conductor de tipo *p* debido al exceso de oxígeno en su estructura, el cual provoca la generación de huecos y vacancias de níquel [4].

El NiO puede ser usado como persiana inteligente puesto que dependiendo de la cantidad de O en exceso puede ser transparente u opaco [5,6]. Es más, su resistividad eléctrica es muy dependiente de dicho exceso, por lo que puede ser aplicado como detector de gas [3]. También este material encuentra un uso en la construcción de celdas fotovoltaicas, actuando como capa búfer [7], además de su potencial aplicación en transistores de nueva generación [8,9] al presentar efectos de conmutación resistiva [8,10,11].

En el presente trabajo se depositaron películas delgadas de NiO mediante la técnica de sputtering por magnetrón en modo reactivo [12] y se analizaron los cambios morfológicos y de resistividad eléctrica al alterar el valor de polarización del sustrato. Dichos análisis intentan aclarar un poco más el comportamiento de este compuesto cuando se somete a campos eléctricos inesperados en su aplicación como transistor o como parte de una celda solar.

## II. MÉTODO

### A. Protocolo de deposición

La deposición se llevó a cabo en un reactor de acero de 9,4 L de capacidad. Se empleó un magnetrón con blanco de níquel (99,95% de pureza) de 20,27 cm$^2$ de superficie. El vacío base alcanzado fue de $10^{-7}$ Torr y la limpieza del blanco se hizo con una descarga de Ar durante 5 min y un pre-sputtering de otros 5 min en las condiciones de deposición. Durante este proceso el sustrato fue aislado con un obturador. La distancia blanco-sustrato fue de 3 cm.

Se usó una mezcla de 85% de Ar y 15% de $O_2$ y cada deposición duró 10 min. Dicha mezcla aseguró que el régimen de trabajo del magnetrón quedase dentro de la zona de transición entre el régimen metálico y el envenenado [13]. La potencia de descarga en modo DC se fijó a 100 W. Los sustratos de vidrio (ERIE Scientific) fueron limpiados con nitrógeno seco para eliminar impurezas y fueron insertados en la cámara de deposición mediante una cámara estanco. La polarización se llevó a cabo con tensión negativa DC de 0, 50, 100, 200, 300, 400 y 500 V (TECHNIX High Voltage Power Supply). La corriente estuvo entre 1-10 mA, incrementándose ligeramente al aumentar la tensión.

### B. Técnicas de caracterización

El espesor de las películas delgadas se determinó mediante profilometría con un dispositivo DEKTAK 8 recurriendo a una doble medida sobre un escalón. La carga del profilómetro fue de 29,43  N y se aplicó el modo de contacto. Los valores consignados corresponden a un promedio de ocho mediciones.

La estructura cristalina quedó determinada mediante difracción de rayos X. Se usó un equipo SIEMENS Diffraktometer D5000 con ánodo de Cu K  a 40 kV y 40 mA. Las ranuras Soller tuvieron una apertura vertical de 1 mm. El tamaño de paso fue de 0,03° y el tiempo de medición 1 s. La configuración empleada fue  /2  entre 20 y 90°.

La composición elemental se obtuvo mediante EDS (siglas en inglés de espectroscopia de rayos X de dispersión energética) en un dispositivo Jeol JSM 5800LV. La tensión del filamento de emisión fue de 5 kV con una corriente de 1,307 nA y un tiempo de adquisición de 60 s.

La resistividad eléctrica se determinó empleando el método de van der Pauw con una configuración de cuatro puntas en un aparato KEITHLEY 236 [12,14]. Se pintaron con Ag los cuatro electrodos y las conexiones se hicieron con hilos de Au de 40  m de diámetro. Dentro de una atmósfera de He pudieron medirse los valores de resistividad hasta una temperatura de 150 K. Ha de mencionarse que la resistividad del vidrio es de 20 P  cm, mientras que la del NiO estequiométrico es de 180 T  cm.

### III. ANÁLISIS DE RESULTADOS

### A. Análisis de profilometría

Se puede observar en la Tabla I que un incremento de tensión de polarización se traduce en una disminución de espesor en la película delgada. Esto puede ser explicado por el efecto de un aumento de bombardeo iónico hacia la película por parte de las cargas positivas que existen en el plasma de descarga [15]. Si bien el número de cargas es muy inferior al número de neutros bajo estas condiciones de deposición [16] la energía cinética que ganan por la vaina de polarización no puede ser despreciada.

TABLA I
ESPESOR DE LAS PELÍCULAS DELGADAS DE NiO ANTE DIFERENTES VALORES DE TENSIÓN DE POLARIZACIÓN DC. EL GUION INDICA UNA MEDIDA DE ESPESOR POCO CONFIABLE DEBIDO A LA GRAN IRREGULARIDAD SUPERFICIAL

| Polarización (V) | Espesor (nm) |
|---|---|
| 0 | 1171,70 |
| 50 | 1690,46 |
| 100 | 1611,92 |
| 200 | - |
| 300 | - |
| 400 | 1171,27 |
| 500 | 892,17 |

Este efecto de menos espesor debido al resputtering del bombardeo iónico y que este sea más acusado a mayores valores de polarización indica que incluso usando sustratos aislantes los campos eléctricos siguen jugando un papel relevante.

### B. Estructura cristalina

En la Figura 1 se pueden identificar las láminas delgadas de NiO depositadas a diferentes valores de tensión de polarización DC. Los principales picos de la estructura pseudo-fcc están claramente presentes en los difractogramas, indicando que la polarización no ha conllevado la amorfización general de las películas delgadas y se conserva su estructura cristalina típica.

Por otro lado, sí se presenta un desplazamiento de los picos de difracción ante el aumento de tensión de polarización [13], indicando la presencia de deformaciones en dicha estructura cristalina y desarrollando, por tanto, tensiones residuales.

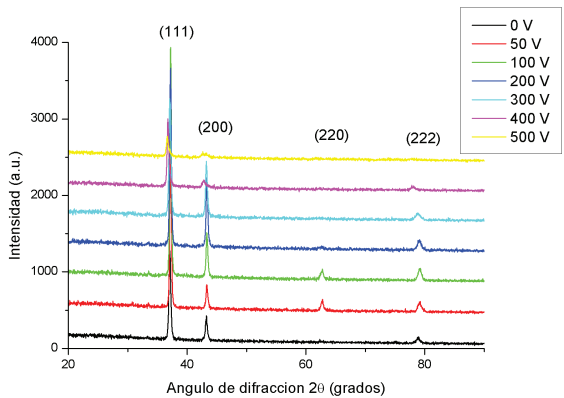

Figura. 1. Difractograma de rayos X en modo Bragg-Brentano que muestra la estructura pseudo-fcc de las películas de NiO depositadas a diferente valor de polarización.

La intensidad de los picos de difracción se puede determinar ajustando funciones voigtianas. Además, se pueden normalizar con respecto los valores de intensidad para la difracción en polvo para determinar la evolución de la textura con el aumento de polarización [17]. En la Figura 2 se muestra la evolución para las presentes muestras tras el cotejado con la ficha de PCPDFWIN vinculada al óxido de níquel (#652901).

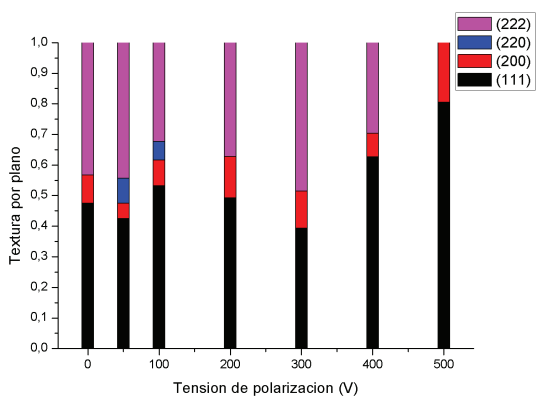

Figura. 2. Contribución de cada plano a la textura global tras la aplicación de la ecuación propuesta en [17].

Se puede observar que, respecto al difractograma patrón, la intensidad de los picos (111) y (222) aumentó en gran manera. El plano (222) solo es una reflexión del (111) [18], por lo que se puede afirmar que dicha textura permanece invariable a la aplicación de tensión de polarización. Sí se advierte que para tensiones de 50 y 100 V se apreciaba una pequeña contribución del plano (220) pero este fue perdido al aumento de polarización en favor de un aumento de intensidad del plano (200). Esta evolución de la textura en función de la polarización puede señalar a la creación de un campo eléctrico por la distribución de carga no compensada sobre el sustrato aislante. O sea, ante los procesos de caracterización llevados a cabo se puede determinar el efecto del campo eléctrico en la evolución de la textura.

### C. Tensiones residuales

Si suponemos que los iones de oxígeno y níquel colisionan con la película durante su crecimiento han de localizarse dentro de la red cristalina. Esto provoca su distorsión y la aparición de tensiones residuales [19]. Es más, los fenómenos de resputtering (la eyección no deseada de material de la película por bombardeo) también pueden alterar dicho valor de tensión residual.

El método de Nelson-Riley permite promediar un parámetro de red con más fiabilidad [20]. Este se basa en representar como eje dependiente los parámetros de red calculados por la ley de Bragg para cada pico y en el eje dependiente cot $\cos$. El valor a elegir será la intersección de la recta de ajuste de puntos con el eje dependiente. Con estos valores se puede estimar la tensión residual a partir de la fórmula propuesta por Mallikarjuna-Reddy *et al.* [21]:

$$\sigma = -\frac{E}{2v}\left(\frac{a - a_0}{a_0}\right), \qquad (1)$$

con $\sigma$ la tensión residual, $E = 200$ GPa el módulo de Young del NiO y $v = 0{,}31$ la razón de Poisson para el NiO. Asimismo, $a$ es el parámetro de red calculado y $a_0 = 4{,}194$ Å es el parámetro de red del NiO sin tensiones. En la Tabla II se consignan los valores de los parámetros de red calculados para cada polarización y su tensión residual asociada.

| Polarización (V) | Parámetro de red (Å) | (-GPa) |
|---|---|---|
| 0 | 4,2018 | 0,60 |
| 50 | 4,2810 | 6,55 |
| 100 | 4,2804 | 6,51 |
| 200 | 4,1929 | -0,08 |
| 300 | 4,2028 | 0,68 |
| 400 | 4,2448 | 3,86 |
| 500 | 4,2324 | 2,92 |

A la vista de estos resultados se puede indicar que la mayoría de los casos mostró una tensión residual de tipo compresivo (excepto para 200 V). La colisión energética del ion con la capa de NiO pudo provocar reordenamiento cristalino y la destrucción de las primeras capas atómicas. La conjunción de ambos factores promueve la entrada de iones de O y Ni hacia zonas más profundas que en el caso de potencial flotante [8,15,22].

La Tabla II indica que al aumentar la tensión de polarización la tensión residual obtenida es mayor. Sin embargo, a 200 V su tensión decae hasta casi ser cero, como si la estructura cristalina de la película se hubiese reordenado para acomodar una gran cantidad de defectos, aclarando por tanto la aparición de gran rugosidad en la superficie. La idea de reordenamiento puede confirmarse con la Figura 2, ya que la contribución del plano (111) a la textura cambia su tendencia al alza cuando la película se depositó a 200 V de polarización. Tras este valor se vuelve a comprobar un aumento progresivo de tensión residual hasta la condición compresiva. En las deposiciones a 300, 400 y 500 V ya no se observa un aporte y a la textura del plano (220) y a 500 V la señal del plano (222) desaparece por completo. Puede ser que tras el reordenamiento la llegada de iones con aún más energía cinética vuelve a deformar la estructura cristalina hasta un valor límite, donde ya la tensión residual vuelve a caer, quizás debido a un progresivo paso hacia la amorfización global de la película delgada. Esto se puede intuir, ya que en la Figura 1, las intensidades de los picos de la capa de NiO depositada a 500 V de polarización (trazo amarillo) muestran los valores más bajos de todos los casos observados.

### D. Composición elemental

En función de lo anteriormente observado se recoge la información de composición elemental de las capas de NiO depositadas a diferente tensión de polarización. Es necesario mencionar que el dispositivo utilizado no tiene buena sensibilidad para la detección del oxígeno, luego los datos consignados en la Tabla III han de interpretarse de manera cualitativa.

<div align="center">

TABLA III

COMPOSICIÓN ELEMENTAL DE ALGUNAS PELÍCULAS DE NiO DEPOSITADAS A DIFERENTE TENSIÓN DE POLARIZACIÓN

</div>

| Polarización (V) | O (%) | Ni (%) |
|---|---|---|
| 100 | 47,64 | 52,36 |
| 200 | 46,73 | 53,27 |
| 300 | 47,10 | 52,90 |
| 400 | 47,79 | 52,21 |

A la vista de los resultados que aparecen en la Tabla III todas las muestras presentan un exceso de níquel en su composición y esta apenas se ve alterada por la aplicación de una u otra tensión de polarización durante su crecimiento. Dichos porcentajes no son muy diferentes a los que se encuentran en las películas de NiO depositadas en el régimen de sputtering de transición [21].

*E. Medidas de resistividad*

Mediante la configuración de cuatro puntas se pudo medir la resistividad de las películas delgadas de NiO a temperatura ambiente [8]. Debido a la proporción de gases usada (85% de Ar y 15% de $O_2$) el régimen de sputtering es el de transición, luego se espera que la cantidad de vacancias de níquel presentes en la película sea muy baja [4]. De esta manera, si se detectan cambios de resistividad cuando la capa se depositó a una tensión de polarización diferente a la flotante se puede demostrar que los efectos del campo eléctrico impuesto en el sustrato son relevantes, a pesar de la naturaleza aislante del vidrio [23,24].

La Tabla IV muestra los valores de resistividad medidos para las películas de NiO depositadas a 100, 300, 400 y 500 V. Se puede apreciar con claridad que en el primer caso (100 V) la naturaleza de la película es aislante, luego el bombardeo iónico no es lo suficientemente energético como para alterar el comportamiento típico de semiconductor. Sin embargo, cuando la tensión de polarización aumentó se aprecia la disminución de resistividad entre 4 y 5 órdenes de magnitud con respecto al caso de 100 V. Este efecto puede indicar la inclusión de O intersticial en la estructura cristalina, provocando que se desencadenen los fenómenos de transporte eléctrico mediante huecos, con el consecuente desarrollo de vacancias de Ni. La fluctuación de valores de resistividad a polarizaciones de 300, 400 y 500 V debe vincularse con la capacidad de la lámina delgada de reacomodarse al bombardeo energético y el desarrollo de tensiones residuales [25]. Por último, aunque EDS indique que las láminas han de ser deficitarias en oxígeno los datos de resistividad desmienten esto. Queda para futuros trabajos discernir esta incertidumbre aunque preliminarmente los autores se inclinan a tener en cuenta la baja sensibilidad del dispositivo de caracterización EDS a la hora de detectar la cantidad de O presente en la lámina delgada.

<div align="center">

TABLA IV

VALORES DE RESISTIVIDAD MEDIDA PARA DIFERENTES CONDICIONES DE TENSIÓN DE POLARIZACIÓN DURANTE LA DEPOSICIÓN DE LA PELÍCULA DELGADA

</div>

| Polarización (V) | Resistividad ( cm) |
|---|---|
| 100 | $12{,}5{\cdot}10^6$ |
| 300 | 380 |
| 400 | $1{,}08{\cdot}10^3$ |
| 500 | 82 |

Por último, se investigó la evolución de la resistividad eléctrica desde la temperatura ambiente hasta una temperatura de 150 K. La Figura 3 muestra una representación tipo Arrhenius de la lámina delgada de NiO depositada bajo una polarización negativa de 300 V. Se puede observar que ante la disminución de temperatura la resistividad comienza a aumentar, fenómeno típico de los procesos térmicamente activados.

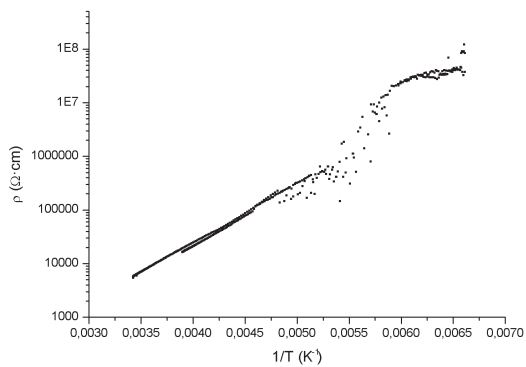

Figura. 3. Medición de la resistividad para la lámina de NiO depositada a 300 V de polarización con el método de van der Pauw en la configuración de cuatro puntas. La representación es de tipo Arrhenius (resistividad *vs* recíproco de la temperatura absoluta). La variación se da desde temperatura ambiente hasta 150 K (ida y vuelta). Se puede observar un abrupto escalón a 167 K. Note la escala logarítmica para el eje de resistividad.

También se puede detectar un aumento abrupto de resistividad cuando se llegó a una temperatura de 167 K. Este gran escalón se debe a un cambio de proceso de conducción eléctrica, cuyo umbral para activarse está justamente a esa temperatura de 167 K [26]. Es más, dicho umbral podría incluso revelar el comportamiento tipo Mott-Hubbard [9] en nuestra muestra, lo que podría ser de interés en la investigación centrada en la conmutación resistiva.

Finalmente, al aumentar la temperatura desde 150 K hasta cerca de la temperatura ambiente no se observó ninguna curva de histéresis, lo que verifica que se recuperan completamente los mecanismos típicos de los semiconductores. Por otro lado, al representar ln( ) frente a $T^{-1}$ se puede medir la pendiente de la recta para determinar así la energía de activación de semiconductor. Para el presente caso $E_a = 0,22$ eV, valor muy bajo en comparación con el del NiO estequiométrico y puro, $E_{NiO} = 3,8$ eV. Esto demuestra que la tensión de polarización aplicada a la capa de NiO durante su crecimiento afecta el nivel de Fermi mediante la inserción de estados dentro de la banda prohibida [27]. Esto también podría presentar interés en la industria solar y de transistores, por ejemplo.

## IV. CONCLUSIONES

En el presente trabajo se depositaron películas delgadas de NiO mediante la técnica de sputtering por magnetrón DC en modo reactivo. Se estudió el efecto de la polarización negativa DC durante el crecimiento de dichas películas. La caracterización mediante profilometría, difracción de rayos X, EDS y resistividad demostraron que si bien el sustrato fue aislante no se puede descartar la influencia de la polarización en las características de dichas películas.

Las películas de NiO perdieron espesor al aumentar la tensión de polarización debido a fenómenos de resputtering por impacto energético. Dicho bombardeo fomentó progresivamente la textura en el plano (111), mientras que las tensiones residuales aumentaron de manera compresiva hasta el valor de polarización de 200 V, momento en que parece reacomodarse la estructura para poder seguir tensionándose hasta los indicios de amorfización a 500 V de polarización. La composición elemental solo pudo conocerse de manera cualitativa, mostrando porcentajes similares de Ni y O. Finalmente, las medidas de resistividad eléctrica a temperatura ambiente demuestran que a altos niveles de polarización la conducción de tipo $p$ es fomentada por la creación de vacancias de Ni. Al bajar la temperatura se demuestra el típico comportamiento de semiconductor y un proceso de activación térmica a 167 K. Se calculó una energía de activación de 0,22 eV, muy inferior al valor establecido.

## AGRADECIMIENTOS

## REFERENCIAS